\documentclass[11pt]{article}

\usepackage[final]{acl}

\usepackage{times}
\usepackage{latexsym}

\usepackage[T1]{fontenc}

\usepackage[utf8]{inputenc}

\usepackage{microtype}

\usepackage{inconsolata}

\usepackage{graphicx}

\usepackage{algorithmic}
\usepackage{graphicx}
\usepackage{textcomp}
\usepackage{colortbl}
\usepackage{xcolor}
\usepackage{multirow} 
\usepackage{booktabs} 
\usepackage{hyperref}
\usepackage{algorithm}
\usepackage{tikz}
\usepackage{threeparttable}
\usepackage{enumitem}
\usepackage{fmtcount}
\usepackage{latexsym}
\usepackage{url}
\usepackage{arydshln}

\usepackage{newfloat}
\usepackage{listings}
\usepackage{verbatim}

\usepackage{booktabs} 
\usepackage{multirow}
\usepackage{tabularx}
\usepackage{makecell}
\usepackage{diagbox}
\usepackage{float}
\usepackage{subcaption}
\usepackage{xspace}
\usepackage{xcolor}
\usepackage{pifont}
\usepackage{balance}
\usepackage{lineno}

\def\BibTeX{{\rm B\kern-.05em{\sc i\kern-.025em b}\kern-.08em
    T\kern-.1667em\lower.7ex\hbox{E}\kern-.125emX}}

\usepackage{adjustbox} 
\usepackage{siunitx}

\usepackage{listings}  
\usepackage{bbding}

\definecolor{codegreen}{rgb}{0,0.6,0}
\definecolor{codegray}{rgb}{0.5,0.5,0.5}
\definecolor{codepurple}{rgb}{0.58,0,0.82}
\definecolor{backcolour}{rgb}{0.95,0.95,0.92}
\definecolor{tblblue}{RGB}{194,218,227} 
\definecolor{hlYellow}{RGB}{255,242,204}

\lstdefinestyle{mystyle}{
    commentstyle= \color{red!50!green!50!blue!50},  
    keywordstyle= \color{blue!70},  
    numberstyle=\tiny\color{codegray},  
    stringstyle=\color{codepurple},
    basicstyle=\ttfamily\footnotesize,
    breakatwhitespace=false,
    breaklines=true,  
    captionpos=b,
    keepspaces=true,
    numbers=left,  
    numbersep=5pt,
    showspaces=false,
    showstringspaces=false,  
    showtabs=false,
    tabsize=2,
    frame=single  
}

\usepackage[most]{tcolorbox}

\newcommand{\figmargin}{\vspace{-3pt}}
\newcommand{\tabmargin}{\vspace{-1pt}}

\newcommand{\boxmargin}{2mm}
\tcbuselibrary{skins, breakable, theorems}
\newtcolorbox{myboxa}[2][]{
    colback=gray!10!white,
    colframe=black, enhanced,
    attach boxed title to top left={yshift=-2mm,xshift=5mm},
    title=#2,#1
}
\newtcolorbox{myboxb}[2][]{
    boxsep=1pt,
    left = \boxmargin, right = \boxmargin, top = \boxmargin, bottom = \boxmargin,
    title={#2},#1
}

\newtcolorbox{myboxc}{
    colback=yellow!10!white,    
    colframe=gray!50,           
    arc = 0pt, outer arc = 5pt,
    boxsep=0pt, left = 3pt, right = 0pt, top = 0pt, bottom = 0pt, 
    leftrule=3pt,               
    bottomrule=0pt, toprule=0pt, rightrule=0pt,
    left = \boxmargin, right = \boxmargin, top = \boxmargin, bottom = \boxmargin,
    before skip=5pt,   
    after skip=5pt    
}

\newcommand{\myauthornote}[3]{}

\title{From Conversation to Evaluation: Benchmarking LLMs on Development Knowledge via SimpleDevQA}

\author{
 \textbf{Jing Zhang\textsuperscript{1,\footnotemark[1]}},
 \textbf{Lianghong Guo\textsuperscript{1,\footnotemark[1]}},
 \textbf{Yanlin Wang\textsuperscript{1,\footnotemark[2]}},
 \textbf{Terry Yue Zhuo\textsuperscript{3}},
\\
 \textbf{Yong Wang},
 \textbf{Mingwei Liu\textsuperscript{1}},
 \textbf{Jiachi Chen\textsuperscript{5}},
 \textbf{Ensheng Shi \textsuperscript{2}},
\\
 \textbf{Yuchi Ma\textsuperscript{2}},
 \textbf{Hongyu Zhang\textsuperscript{4}},
 \textbf{Zibin Zheng\textsuperscript{1}}
\\
 \textsuperscript{1}Sun Yat-Sen University, Zhuhai Key Laboratory of Trusted Large Language Models, 
 \\
 \textsuperscript{2}Huawei CodeArts Model Team, 
 \textsuperscript{3}Monash University,
 \textsuperscript{4}Chongqing University, 
 \\
 \textsuperscript{5}The State Key Laboratory of Blockchain and Data Security, Zhejiang University
\\
 \small{
   \href{mailto:email@domain}{https://github.com/DeepSoftwareAnalytics/SimpleDevQA}
 }
}

\begin{document}
\maketitle

\begin{abstract}
The Development Knowledge Question Answering (Dev Knowledge QA) task aims to provide accurate natural language answers to knowledge-seeking questions during software development.
To investigate the importance of Dev Knowledge QA in AI-assisted software development and to what extent it has been explored, we conduct a preliminary analysis on real user–LLM dialogues from WildChat. Our findings indicate that Dev Knowledge QA plays a significant role in real-world software development scenarios, and these raw dialogues cannot be directly used to construct a Dev Knowledge QA benchmark. 
Existing Dev Knowledge QA benchmarks are limited in development knowledge scope and often not built from real user queries.
To bridge this gap, we design a three-phase pipeline that transforms real-world dialogue into simple development knowledge-seeking QA pairs.
Through this pipeline, we introduce SimpleDevQA, a multilingual Dev Knowledge QA benchmark inspired by real user dialogues. 
This dataset covers three languages (English, Chinese, and Russian), and focuses on questions with unique, short, and verifiable answers, making evaluation more accurate and simple.
Extensive experiments with 21 mainstream LLMs show that closed-source models generally perform best on SimpleDevQA. We also find that RAG-based knowledge injection improves accuracy, and that Dev Knowledge QA performance correlates with both model confidence and code-generation capability.
To facilitate the replication study, we have released our data and code at:~\url{https://github.com/DeepSoftwareAnalytics/SimpleDevQA}.
\end{abstract}

\renewcommand{\thefootnote}{\fnsymbol{footnote}} 
\footnotetext[1]{Jing Zhang and Lianghong Guo have equal contribution.}
\footnotetext[2]{Yanlin Wang is the corresponding author.}

\section{Introduction}

\begin{figure}[t]
    \centering
    \includegraphics[width=1\columnwidth]
    {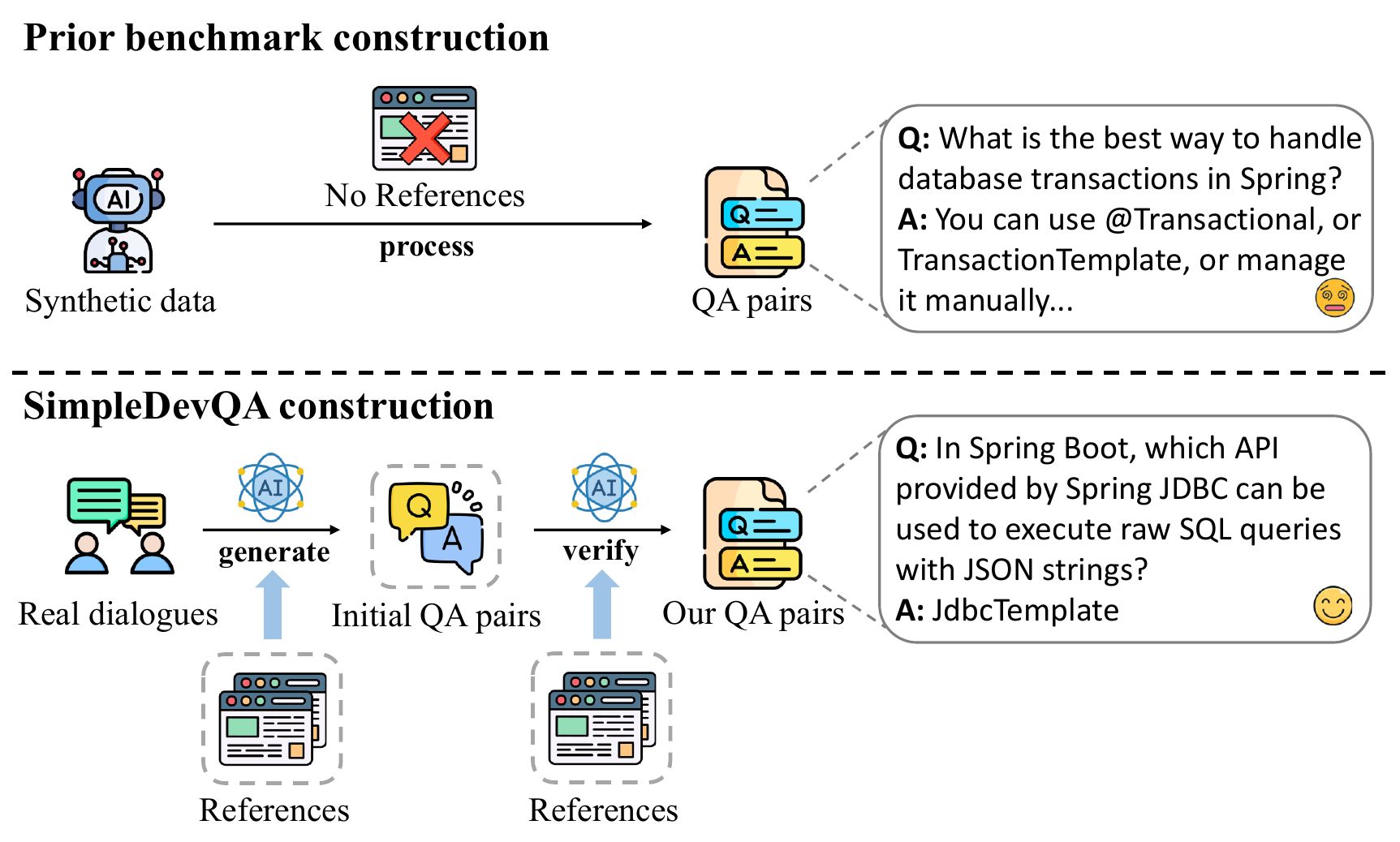} 
    \caption{
    \textbf{SimpleDevQA vs. prior Dev Knowledge QA benchmarks.
    } Synthetic QA pairs (top) lacks references and may admit incorrect answers, making automated evaluation unreliable. Real dialogue-derived QA pairs produced by SimpleDevQA (bottom) has a single verifiable gold answer, supported by references, enabling reliable and accurate evaluation.}
    \label{fig:case} 
\end{figure}

The Development Knowledge Question Answering (Dev Knowledge QA) task focuses on answering knowledge-based questions that arise during software development.
Effectively responding to these diverse, knowledge-seeking questions is crucial for boosting developer productivity and enhancing software quality. With the increasing integration of LLMs into developer workflows, Dev Knowledge QA has emerged as a critical area for research and evaluation~\cite{da2020short, zheng2025towards, 11052799}.

To investigate the importance of Dev Knowledge QA in AI‐assisted software development scenarios, we conduct a preliminary study on WildChat~\cite{zhao2024wildchat}, a dataset including one million user–ChatGPT~\cite{welsby2023chatgpt} conversations. 
Considering the large scale and topic diversity of WildChat, 
we filter and sample it to obtain WildChat-Dev-Lite, a software development-related subset of WildChat.
We obtain three findings based on the preliminary study on the WildChat-Dev-Lite: 
\begin{itemize}[nosep]
    \item The Dev Knowledge QA task accounts for the highest proportion (39.6\%) of all tasks observed during user interactions with LLMs.
    \item In real Dev Knowledge QA dialogues, only 27.5\% of dialogues focus on code understanding. 
    \item Only 17.1\% of real Dev Knowledge QA dialogues could be used for constructing a benchmark.
\end{itemize}

Through the preliminary study, we identify the importance of constructing a Dev Knowledge QA benchmark. 
However, existing Dev Knowledge QA benchmarks have the following two key problems:

\textbf{P1: Limited development knowledge scope.} 
Existing Dev Knowledge QA benchmarks mainly focus on code understanding~\cite{liu2021codeqa, lee2022cs1qa, chen2025coreqauncoveringpotentialslanguage, hu2024coderepoqa, liu2024repoqa, NEURIPS2024_e888eb94} and do not cover other knowledge related to the development process.
For instance, CS1QA~\cite{lee2022cs1qa} primarily evaluates code understanding in programming education scenarios. However, 
based on Section~\ref{sec:psrq2} from our preliminary study, practical software development demands a much broader knowledge.

\textbf{P2: Not built from real user queries.}  
Some existing Dev Knowledge QA benchmarks~\cite{liu2021codeqa, liu2024repoqa} are not built from real user queries from real-world development scenarios, which can not reflect the real demands of developers (Figure~\ref{fig:case} top).
For example, RepoQA~\cite{liu2024repoqa} focuses only on long-context tasks from code repositories, which fails to fully reflect genuine developer task demands in real development scenarios.

In this paper, we propose SimpleDevQA, a Dev Knowledge QA benchmark built from real developer dialogues, to address the aforementioned problems. 
Based on the findings in our preliminary study, we design and implement a three-phase data construction pipeline to convert real-world SE-related conversations into a Dev Knowledge QA benchmark. 
When constructing the benchmark, we focus only on software development knowledge-seeking questions that have a single correct answer, which helps us avoid the open-endedness of language models, following prior work~\cite{wei2024measuring,he2024chinese}. This is important because it makes measuring factuality much easier, though it may leave out questions that could have multiple valid responses~\cite{zhang2025siren, guan2024deliberative}. 
Additionally, each QA pair is also accompanied by multiple web-retrieved references, which can be used to verify the factual accuracy of the answer (Figure~\ref{fig:case} down). 

We summarize the main contributions as follows:
\begin{itemize}[nosep]
   \item We conduct a preliminary study to assess the importance of Dev Knowledge QA in AI-assisted software development.
   
   \item We propose SimpleDevQA, a multilingual Dev Knowledge QA benchmark inspired by real user dialogues.
  
   \item We construct a pipeline that converts real-world conversations into a Dev Knowledge QA benchmark.

   \item We conduct extensive experiments on SimpleDevQA and find that the performance of LLMs in Dev Knowledge QA.

\end{itemize}

\section{Related Work}






\subsection{Dev Knowledge QA Benchmarks}

Numerous Dev Knowledge QA benchmarks have emerged within the field of software engineering benchmark.
s~\cite{liu2021codeqa, lee2022cs1qa, chen2025coreqauncoveringpotentialslanguage, hu2024coderepoqa, liu2024repoqa, NEURIPS2024_e888eb94, li2024infibench}.

Existing Dev Knowledge QA benchmarks rely on questions grounded in a given code snippet, while overlooking broader development knowledge~\cite{lee2022cs1qa, NEURIPS2024_e888eb94,hu2024coderepoqa}.
For example, CodeQA~\cite{liu2021codeqa} targets snippet-level syntax/API understanding, 
CoReQA~\cite{chen2025coreqauncoveringpotentialslanguage} uses code-review history from GitHub issues/comments, 
RepoQA~\cite{liu2024repoqa} evaluates repository-level multi-turn and long-context understanding, and InfiBench~\cite{li2024infibench} scales to free-form QA from high-quality Stack Overflow questions.
The comparison between SimpleDevQA and other Dev Knowledge QA benchmarks is shown in Table~\ref{tabbenchmark}.

\subsection{Real-World Software QA Datasets} 
Numerous software QA datasets have been constructed from online QA communities~\cite{wang2022execution,dhingra2017quasar, wu2023retrieving}.
For example, the StaQC~\cite{yao2018staqc} focuses on matching natural language questions with relevant code snippets, 
the TechQA corpus~\cite{castelli2019techqadataset} consists of 
questions from a technical forum. 
However, these real QA datasets often yield verbose answers, contain subjective opinions, or present multiple solutions, making it difficult to directly assess LLMs' performance.

\subsection{Factuality Benchmarks} 
Factual evaluation of LLMs has garnered academic attention, leading to the development of several representative benchmarks
~\cite{chern2023factool,zhong-etal-2024-agieval,huang2023c,srivastava2022beyond,zhang2025llm}. 
For example, TruthfulQA~\cite{lin2021truthfulqa} assesses whether models generate answers that sound reasonable but are actually false. HallEval~\cite{li2023halueval} measures hallucination across QA, summarization, and dialogue tasks. Additionally, OpenAI recently released SimpleQA~\cite{wei2024measuring} and Chinese SimpleQA~\cite{he2024chinese}, which enable reliable factual accuracy assessment. 
However, current factual benchmarks remain limited to general domains, leaving professional fields like software development without dedicated evaluation frameworks.

\section{Preliminary Study}
\label{sec:psrq}

To investigate the importance of the Dev Knowledge QA task in real-world development scenarios, we conduct a preliminary study on WildChat~\cite{zhao2024wildchat}.
Given the diverse topics in WildChat's dialogues, we extract software development-related dialogues for our research. For the detailed procedure, we refer our readers to Appendix~\ref{sec:appendix_extract}. 

\subsection{Importance Analysis of Dev Knowledge QA}
\label{sec:distribution}

To investigate the importance of Dev Knowledge QA, we analyze the distribution of Dev Knowledge QA tasks in real-world development scenarios when users utilize LLMs. We manually filter dialogues from WildChat-Dev-Lite that are unrelated to software development and classify the real-world dialogues into 8 task categories. The detailed classification procedure is provided in the Appendix~\ref{sec:appendix_classfy}.
Then, we perform a statistical analysis on the classified dialogues in the Figure~\ref{figPSRQ1}. 

As illustrated in the Figure~\ref{figPSRQ1} below, we find that the Dev Knowledge QA task (39.6\%) accounts for the highest proportion among all tasks, surpassing the code generation task (32.3\%), illustrating the high popularity of the Dev Knowledge QA task in real-world development scenarios.


\subsection{Topic Analysis of Dev Knowledge QA}
\label{sec:psrq2}
Since previous Dev Knowledge QA benchmarks primarily focus on code understanding~\cite{liu2021codeqa, hu2024coderepoqa, lee2022cs1qa, liu2024repoqa}, we analyze whether existing benchmarks can fully evaluate real Dev Knowledge QA capabilities. 

Through manual annotation of 3,098 Dev Knowledge QA dialogues, we find that only 851 (approximately 27.5\%) involve questions centered on specific code snippets.
In contrast, the remaining 2,247 dialogues (approximately 72.5\%) consisted of queries seeking factual information about other aspects of development knowledge, such as system design, database management, and computer architecture, etc. 
Thus, existing Dev Knowledge QA benchmarks can not fully evaluate real Dev Knowledge QA capabilities.


\begin{figure}[t]
    \centering
    \includegraphics[width=\columnwidth]{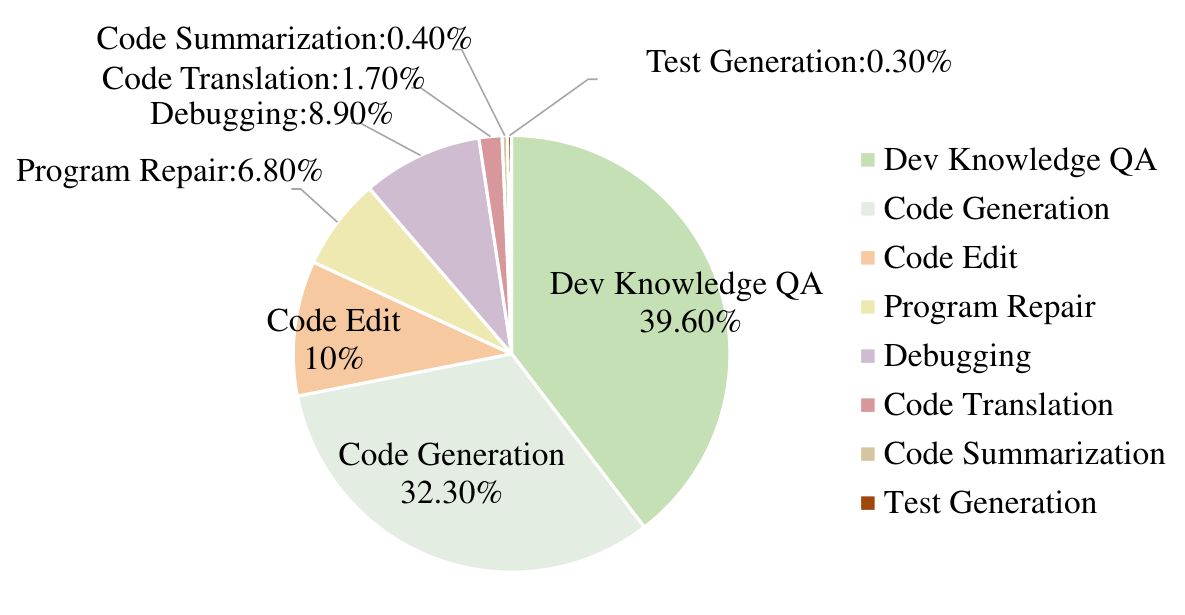} 
    \caption{The distribution of SE tasks in WildChat-Dev-Lite dialogues.}
    
    \label{figPSRQ1} 
\end{figure}

\subsection{Challenge Analysis of Constructing Dev
Knowledge QA Benchmark}
We further investigate whether these real-world development-related dialogues can be used to construct a Dev Knowledge QA benchmark. 

First, We randomly sample 1,000 Dev Knowledge QA dialogues for manual verification. Then, we manually verify each Dev Knowledge QA instance against three critical criteria proposed by previous research~\cite{wei2024measuring, he2024chinese}: (1) the question must have a single answer, (2) reference answers should not change over time, and (3) reference answers must be supported by evidence.
The results show that only 17.11\% of the dialogues are suitable for building a Dev Knowledge QA benchmark. Most dialogues cannot be used because many answers are incorrect, and even when correct, the questions often admit multiple valid responses, making it difficult to define a single reference answer. The cases are shown in Appendix~\ref{sec:appendixcase}. This highlights the need for a dedicated benchmark that can automatically evaluate LLMs on Dev Knowledge QA in real-world development scenarios.


\section{SimpleDevQA}

\begin{figure*}[t]
    \centering
    \includegraphics[width=\linewidth]{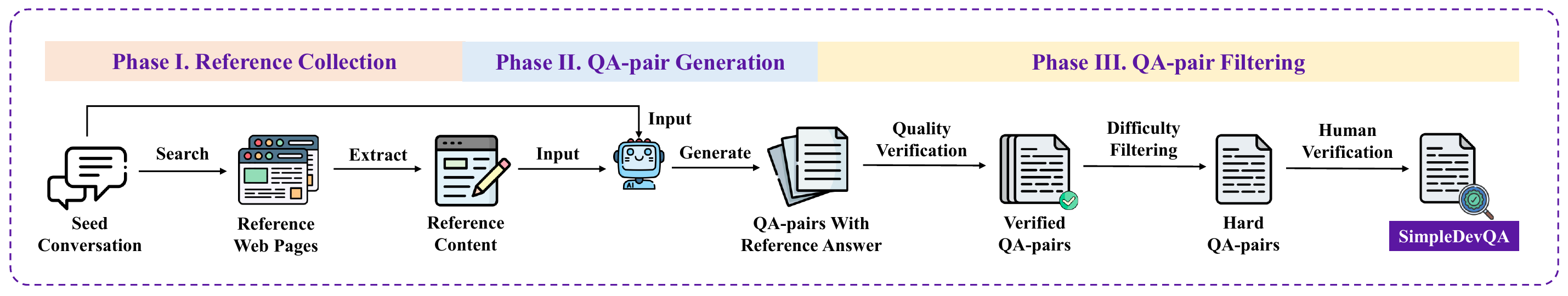} 
    \caption{SimpleDevQA Construction Pipeline.}
    \label{fig:construction_pipeline} 
\end{figure*}


Our preliminary study in the Section~\ref{sec:psrq} shows that most real-world Dev Knowledge QA dialogues are unsuitable for benchmark construction. 
To address this, we design a three-stage benchmark construction framework that can convert real-world SE-related conversations into a Dev Knowledge QA benchmark. An overview of our pipeline is illustrated in Figure~\ref{fig:construction_pipeline}. 
First of all, we randomly sample 3,000 real-world conversations from WildChat-Dev to serve as the seed conversation. 

\subsection{Reference Collection}

To ensure that the generated QA pairs are grounded in factual evidence while remaining faithful to the original user intent, we collect high-quality reference documents for each real-world developer dialogue in this stage.  

Starting from the seed dataset, each dialogue is analyzed by GPT-4o-mini~\cite{openai2024gpt4omini} to identify core software engineering topics and then generate search queries that capture the main intent and technical focus of the dialogue. These queries are submitted to the Google Search API~\cite{piasecki2018google} to retrieve relevant web page URLs. For each dialogue, we collect 10 candidate web pages. 
Then, we use the Goose3~\cite{goose32024} tool to remove irrelevant content from the collected web pages. Finally, we collect 30,000 reference documents for the QA-pair Generation stage.

\subsection{QA-pair Generation}

To ensure both the realism and reliability of the constructed QA pairs, we generate Dev Knowledge QA pairs using the original real-world dialogue and its corresponding reference documents. 

The generation process begins with the preprocessing of the reference content. For documents that are too long to be directly used as input for the LLM due to context window limits or irrelevant information, 
we first extract their main content. 
Then, all reference documents linked to the same dialogue are combined into a single reference text. This unified reference text, along with the original real-world dialogue, is fed into an LLM to produce candidate QA pairs. 
Details on the generation can be found in the Appendix~\ref{sec:appendix1}.
Finally, we generate 9,000 QA pairs from the initial 3,000 Dev Knowledge dialogues.


\subsection{QA-pair Filtering}


\textbf{Quality Verification.} To improve the quality of SimpleDevQA, we separately verify the quality of both questions and answers.
First, to improve the quality of questions, we filter out low-quality questions.
Specifically, following previous studies\cite{he2024chinese}, we use three core standards for QA pair quality verification:
(1) Each question must have exactly one unambiguous and uncontroversial answer, avoiding open-ended questions;
(2) Each question should exhibit sufficient complexity to evaluate LLMs' depth of understanding of software engineering knowledge;
(3) Each question must have verifiable answers based on publicly available information by December 31, 2024.
In this step, we apply LLM to filter out the samples that fail to meet any of these criteria.
Second, to improve the quality of answers, we implement an LLM with Retrieval-Augmented Generation (RAG)~\cite{lewis2020retrieval} system to verify the factual correctness of answers.
For our RAG system, we utilize LlamaIndex~\cite{Liu_LlamaIndex_2022} to build the search tool and select Google Search~\cite{piasecki2018google} as the search engine. This tool is used to search related web pages as references for subsequent answer verification. Then, we use 
our RAG system to classify the QA pairs with incorrect answers. Finally, we manually correct the answers of these QA pairs. After Quality Verification, approximately 6,020 pairs (67\%) were retained, with 33\% discarded. 
 
\textbf{Difficulty Filtering.} 
To increase the difficulty of SimpleDevQA, we employ LLMs to filter out low-difficulty QA pairs. 
Importantly, the notion of difficulty here is \emph{model-centric}: a question is considered difficult if strong LLMs are likely to answer it incorrectly, rather than the \emph{human-centric} notion that a difficult problem merely requires more time or effort to solve. Specifically, following previous research~\cite{he2024chinese}, we use four LLMs with strong general capability: GPT-4o~\cite{openai2024gpt4o}, Llama-3-70B-Instruct~\cite{grattafiori2024llama}, Qwen2.5-72B-Instruct~\cite{qwen5qwen2} and GLM-4-Plus~\cite{glm42024}. If a question can be correctly answered by all four strong models, it is considered easy.
After this step, we obtain 3,231 difficult QA pairs and 2,789 easy QA pairs.

\textbf{Human Verification.} In this stage, we conduct human verification to further enhance the quality of SimpleDevQA. We follow the criteria in Quality Verification to verify data quality. Each QA pair is evaluated independently by two annotators. They first check whether it meets the predefined standards: if either annotator rejects it, the pair is removed. Then, annotators search for supporting evidence using search engines and provide answers with citations from authoritative sources (at least two URLs per answer). If the two annotators disagree, a third annotator reviews the case and makes the final decision, considering both previous evaluations. 
Only QA pairs with full agreement are kept to ensure high precision and consistency with the standards. Finally, we obtain 5,529 high-quality QA pairs.


\section{Benchmark Characteristics}

SimpleDevQA is a Dev Knowledge QA benchmark designed for the software development domain, containing 5,529 manually verified QA pairs. This includes 2,740 difficult QA pairs and 2,789 easy QA pairs. The feature statistics of our dataset are presented in the Appendix~\ref{sec:appendix3}. Additionally, to analyze the development knowledge evaluated in our benchmark, we manually classify the questions into syntactic and semantic categories following a prior study (see the Appendix~\ref{sec:appendix3} for details).

In summary, SimpleDevQA includes the following key characteristics:

\textbf{Built from real user queries.} The questions in SimpleDevQA are built from dialogues in the real-world development scenarios, which can reflect the real task demands of developers.

\textbf{Diverse development knowledge.} 
Based on the classification of the QA pairs, we find that SimpleDevQA covers a broader range of development knowledge domains, which more realistically reflects the development knowledge QA ability of LLMs.

\textbf{Multilingual.} SimpleDevQA incorporates three popular languages, including Chinese, English, and Russian.
This dataset can be used to evaluate the multilingual Dev Knowledge QA capabilities of different LLMs.


\textbf{Static.} In SimpleDevQA, each QA pair's reference answers are grounded in stable development knowledge, which remains invariant over time or external changes. Each QA pair is equipped with reference web documents, which serve as static sources of knowledge. This design ensures long-term reproducibility for model evaluation, 

\textbf{Easy to evaluate.} 
We retain only those questions with unambiguous answers, which ensures that we can evaluate the correctness of predicted answers simply using the LLM-as-judge method~\cite{zhuo2024ice, li2024llms, son2024llm, he2025code}, according to previous studies~\cite{he2024chinese,wei2024measuring}.

\section{Experimental Setup}

\subsection{Experimental Details}

 \textbf{Model Selection.} 
 In the evaluation, we choose 21 widely used LLMs.
  For the closed-source models, we include Claude-3.5-Haiku~\cite{anthropic2024haiku}, o3-mini~\cite{openai2024o3mini}, GPT-3.5-Turbo\cite{brown2020language}, DeepSeek-V3.2~\cite{deepseekai2025deepseekv32pushingfrontieropen}, DeepSeek-V3~\cite{liu2024deepseek}, GPT-4o~\cite{openai2024gpt4o}, and DeepSeek-R1. For the open-source models, 
 we include Qwen3-30B-A3B-Instruct~\cite{yang2025qwen3technicalreport}, the Qwen2.5 series (7B, 14B, 32B)~\cite{qwen5qwen2}, the InternLM2.5 series (7B, 20B)~\cite{team2024internlm2}, the Llama3 series (8B, 70B)~\cite{grattafiori2024llama}, the DeepSeek-Coder series (6.7B, V2-lite)~\cite{guo2024deepseek}, Qwen3-Coder-30B-A3B-Instruct~\cite{yang2025qwen3technicalreport}, the Qwen2.5-Coder series (7B, 32B)~\cite{hui2024qwen2}, and CodeLlama-7B-instruct~\cite{roziere2023code}. 

\textbf{Experiment Settings.} During LLM inference, we set the temperature to 0.7 and the top-p to 0.95. During evaluation, we set the temperature of the judge model to 0.5 and top-p to 1. 
This setting mitigates the risk of extreme bias from greedy decoding while avoiding excessive noise, thereby producing judgments that are both consistent and reasonably diverse~\cite{li2024llms, gu2024survey}. Nevertheless, it may introduce a certain degree of randomness and potential bias.
To mitigate randomness, we conduct each experiment three times and average the outcomes. 

\subsection{Evaluation Metrics}
\textbf{Grading Method:} Following previous studies~\cite{wei2024measuring,he2024chinese}, we evaluate the correctness of the model’s predicted answers by prompting a separate judge model with both the prediction and the reference answer to assign one of three labels: Correct, Not Attempted, or Incorrect. 
We here use GPT-4o-mini~\cite{openai2024gpt4omini} as the judge model. 

\textbf{Metrics:} Following previous studies~\cite{wei2024measuring,he2024chinese}, we use these metrics to evaluate the performance of LLMs on SimpleDevQA:
\begin{itemize}[nosep,leftmargin=10pt]
    \item\textbf{Correct (CO):} The predicted answer includes the reference answer without contradiction.
    \item \textbf{Not Attempted (NA):} 
    The model effectively does not attempt a full answer: either it refuses/admits its inability, or it gives only a partial response that fails to fully match the reference answer, and there are no contradictions with the reference answer.
    \item \textbf{Incorrect (IN):} The predicted answer contradicts the reference answer at any point, even if the contradiction is later corrected.
    \item \textbf{Correct Given Attempted (CGA):} The ratio of correct answers among all attempted answers. 
    \item \textbf{F-score:} 
    The harmonic mean of the overall percentage of correctly answered questions and Correct Given Attempted. 
    \item \textbf{Average Tokens:} The average sum of input and output tokens per question that can evaluate LLM's potential invocation costs.
\end{itemize}

\section{Experimental Results}


    

\subsection{Performance Comparison Analysis}

\renewcommand{\arraystretch}{1.25} 
\begin{table*}[t]
\centering
\small
\tabmargin
\begin{tabular}{lccccccc}
\toprule
\textbf{Model} & \textbf{Type} & \textbf{Correct} & \textbf{Incorrect} & \textbf{NA} & \textbf{CGA} & \textbf{F-score} &  \textbf{Avg. Tokens} \\ 
\midrule
\rowcolor{cyan!15} \multicolumn{8}{c}{Closed-Source Large Language Models} 
\\
Claude-3.5-Haiku                & General LLM & 0.536   & 0.324     & \textbf{0.14}          & 0.623                   & 0.576 & 321.62 \\
o3-mini        & General LLM                  & \textbf{0.718}   & 0.278     & \underline{0.004}        & 0.721                   & \textbf{0.719} & 229.86 \\
GPT-3.5-Turbo      & General LLM              & 0.677   & 0.307     & 0.016         & 0.688                   & 0.682 & 117.05 \\
DeepSeek-V3.2       & General LLM               & 0.6   & 0.387     & 0.013         & 0.608                   & 0.604 & - \\
DeepSeek-V3       & General LLM               & 0.56   & 0.432     & 0.007         & 0.564                   & 0.562 & 527.57 \\
GPT-4o      & General LLM                     & 0.619   & 0.373     & 0.008         & 0.624                   & 0.622 & 254.81 \\
DeepSeek-R1     & General LLM                 & 0.679   & \textbf{0.262}     & 0.059         & \textbf{0.722}                   & 0.7 & \underline{918.02}   \\ 
\rowcolor{cyan!15} \multicolumn{8}{c}{Open-Source Large Language Models}                                                     \\ 
Qwen3-30B-A3B-Instruct    & General LLM         & 0.638   & 0.355     & 0.007         & 0.642                   & 0.64 & - \\
Qwen2.5-32B-Instruct    & General LLM         & 0.543   & 0.442     & 0.015         & 0.551                   & 0.547 & 309.92 \\
Qwen2.5-14B-Instruct   & General LLM          & 0.498   & 0.489     & 0.013         & 0.505                   & 0.501 & 317.04 \\
Qwen2.5-7B-Instruct    & General LLM          & 0.445   & 0.534     & 0.021         & 0.454                   & 0.45 & 360.53  \\
InternLM2.5-7B-chat   & General LLM           & 0.411   & 0.572     & 0.017         & 0.418                   & 0.414 & 521.66 \\
InternLM2.5-20B-chat  & General LLM           & 0.466   & 0.517     & 0.017         & 0.474                   & 0.47 & 458.14 \\
Llama-3.1-8B     & General LLM                & 0.449   & 0.535     & 0.016         & 0.456                   & 0.453 & 304.89 \\
Llama-3.1-70B    & General LLM                & 0.538   & 0.451     & 0.011         & 0.544                   & 0.541 & 304.82 \\
CodeLlama-7B-Instruct    & Code LLM        & \underline{0.389}   & \underline{0.594}     & 0.017         & \underline{0.396}                   & \underline{0.393} & 403.62 \\
DeepSeek-Coder-V2-Lite-Instruct & Code LLM & 0.518   & 0.472     & 0.01          & 0.523                   & 0.521 & 403.87 \\
DeepSeek-Coder-6.7B-Instruct  & Code LLM   & 0.511   & 0.477     & 0.012         & 0.517                   & 0.514 & 269.88 \\
Qwen3-Coder-30B-A3B-Instruct & Code LLM & 0.582   & 0.403     & 0.015          & 0.591                   & 0.586 & - \\
Qwen2.5-Coder-7B-Instruct  & Code LLM      & 0.573   & 0.412     & 0.015         & 0.582                   & 0.578 & \textbf{116.9} \\
Qwen2.5-Coder-32B-Instruct  & Code LLM   & 0.574   & 0.412     & 0.014         & 0.582                   & 0.578 & 278.35 \\
\bottomrule
\end{tabular}
\caption{Results of different models on difficult QA pairs from SimpleDevQA. NA is short for Not Attempted; CGA is short for Correct Given Attempted.}
\label{tab2}
\end{table*}
\renewcommand{\arraystretch}{1}

\begin{figure}[t]`
    \centering
    \includegraphics[width=\columnwidth]{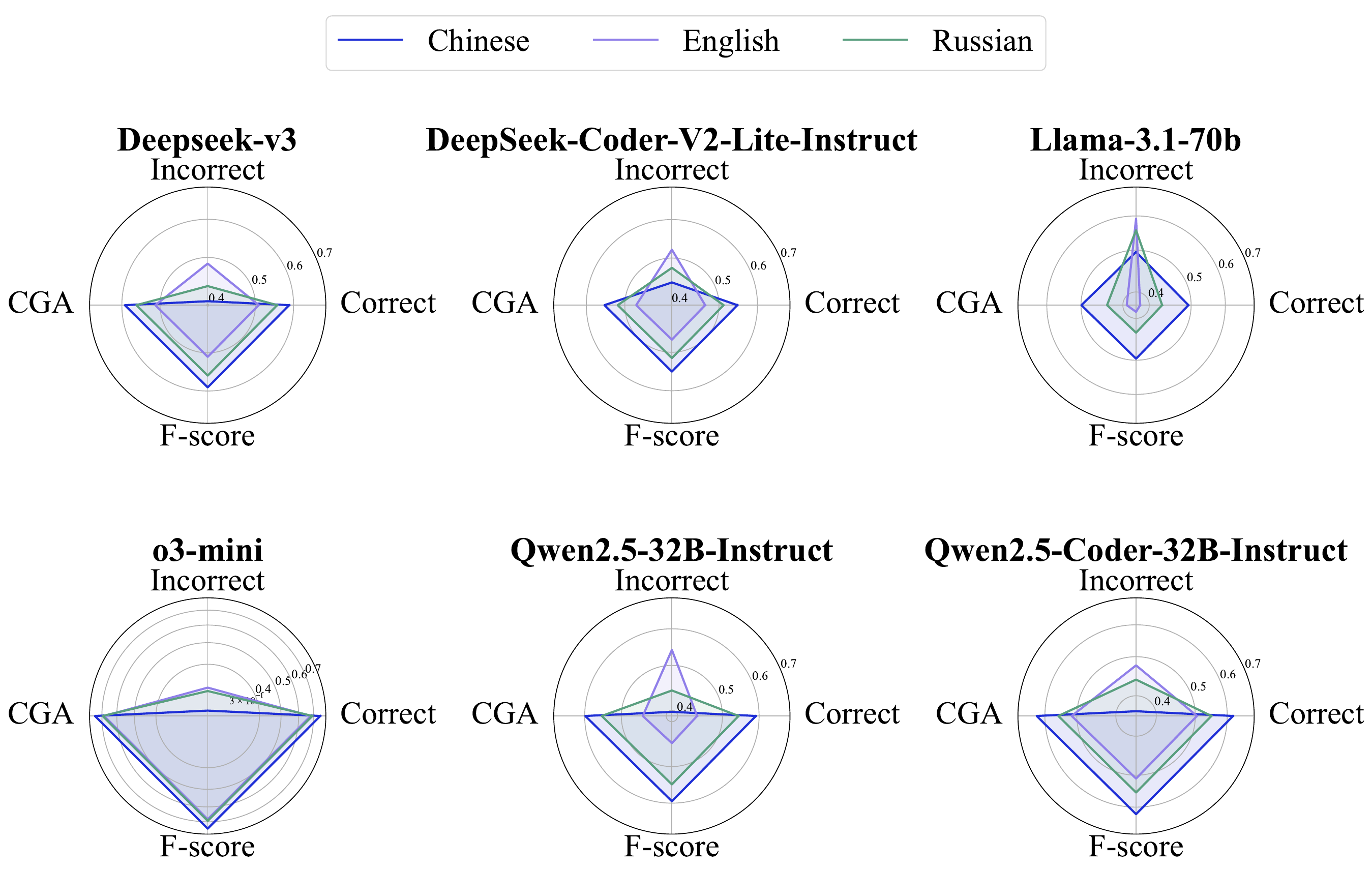} 
    \caption{Results of different models between different language subsets.}
    \label{figlang} 
\end{figure}
We evaluate 21 mainstream LLMs on the difficult QA pairs in SimpleDevQA. The results are presented in the Table~\ref{tab2}.
First, we can find that LLMs' performance on SimpleDevQA varies considerably, and o3-mini and DeepSeek-R1 perform best, achieving F-scores of 0.719 and 0.7, respectively. 
Second, at similar parameter scales, specialized code LLMs significantly outperformed general LLMs. For instance, while Qwen2.5-32B-Instruct scored only 0.547 on the F-score, Qwen2.5-Coder-32B-Instruct achieved 0.578. 
Third, we find that closed-source models generally outperformed open-source models, with closed-source models consistently achieving higher F-scores. For example, the open-source Llama-3.1-70B has an F-score of 0.541, while the closed-source Claude-3.5-Haiku achieved 0.576. 

Furthermore, we find that models' performance improves as their scale increases across many model series, such as the InternLM2.5 and Llama3 series. 
Finally, we find that Claude-3.5-Haiku achieves the highest score on the Not Attempt metric among all evaluated LLMs. For questions about which it is uncertain, it tends to abstain rather than provide an incorrect answer.

We also analyze the performance of six LLMs on the Chinese, English, and Russian subsets of SimpleDevQA, with the results detailed in Figure~\ref{figlang}. From these experimental outcomes, we identify several key findings: First, o3-mini demonstrates the strongest performance, achieving the leading score across all three languages. Furthermore, most Chinese-developed models, such as the Qwen and DeepSeek series, perform best in Chinese. The result indicates that the model performs differently across languages.

\begin{center}
    \begin{myboxc}{\textbf{Summary: }First, LLMs' performance in Dev Knowledge QA varies considerably.
    Second, there is room for growth in the open-source models' understanding of software development knowledge. 
    Third, code-specific LLMs generally outperform general LLMs of similar scale.}
    \end{myboxc}
\end{center}

\subsection{Impacts of the Knowledge Injection Strategy}

\begin{figure}[t]
    \centering
    \includegraphics[width=\columnwidth]{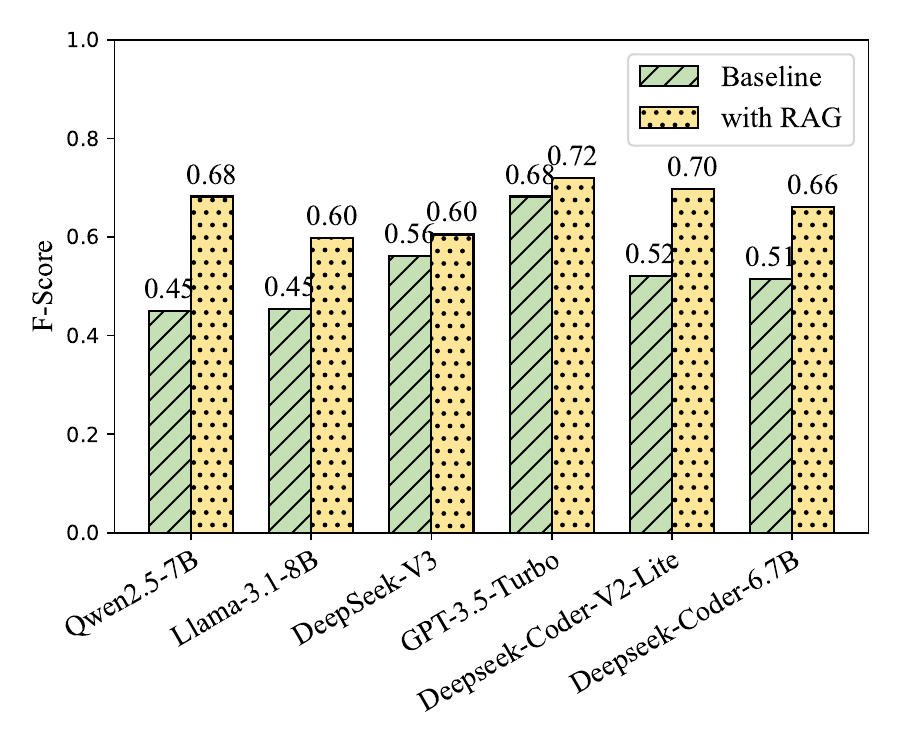} 
    \caption{The effect of the knowledge injection strategy.}
    \label{figRQ2} 
\end{figure}


Previous studies~\cite{xia2024rule,li2024rac,li2024enhancing} have demonstrated that the Retrieval Augmentation Generation~\cite{lewis2020retrieval} (RAG) strategy is one common method to inject factual knowledge into LLMs to improve LLMs' factual accuracy on general QA tasks. In this section, we investigate its impact on LLMs' performances in the Dev Knowledge QA task. 
Following previous studies~\cite{he2024chinese}, we implement our RAG pipeline using LlamaIndex~\cite{Liu_LlamaIndex_2022} and integrate the Google Search API~\cite{googlecustomsearch2024} to retrieve relevant software-engineering knowledge from the web. 

The experimental results are presented in the Figure~\ref{figRQ2}. First, we can find that after implementing the RAG strategy, all evaluated LLMs show significant performance improvements on SimpleDevQA, with an average improvement of 13.2\%. 
Second, the RAG approach effectively reduces the performance gap between different LLMs. For instance, without RAG, the F-score gap between Qwen2.5-7B-Instruct and GPT-3.5-Turbo is as high as 23.2\%, whereas after RAG integration, this gap decreases substantially to just 3.7\%. The results show that smaller LLMs can achieve comparable performance to larger LLMs after the RAG strategy.

\begin{center}
    \begin{myboxc}{\textbf{Summary: }The knowledge injection strategy can improve LLMs' accuracy in Dev Knowledge QA, and can enable smaller LLMs to achieve performance comparable to that of larger LLMs.
    }
    \end{myboxc}
\end{center}

\subsection{Stated Confidence and Accuracy Correlation Analysis}

\begin{figure}[t]
    \centering
    \includegraphics[width=0.9\columnwidth]{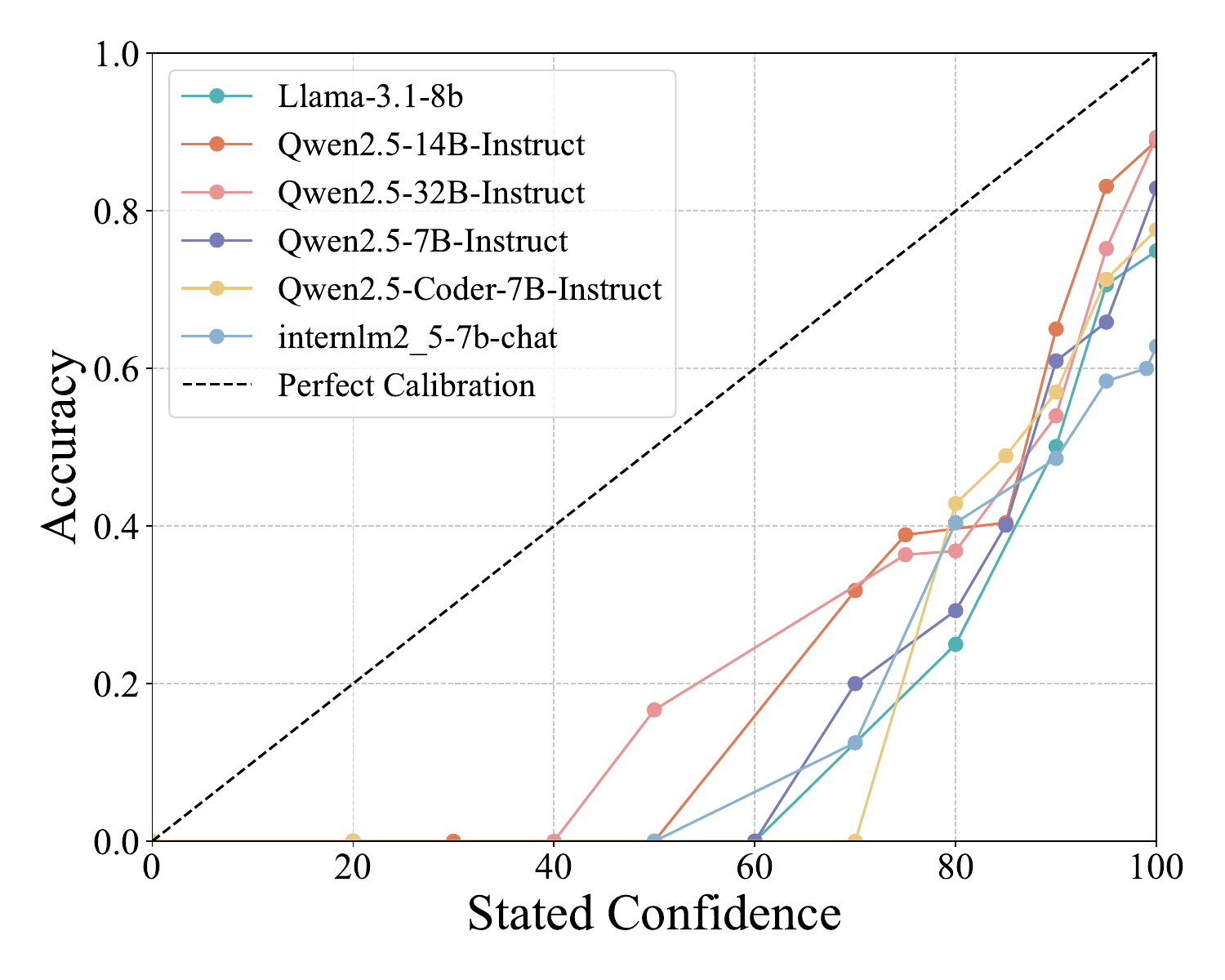}
    \caption{Calibration of model outputs using self-reported stated confidence scores.}
    \label{figRQ3-1}
\end{figure}

According to previous studies~\cite{wei2024measuring,he2024chinese}, the factual QA benchmark like SimpleQA~\cite{wei2024measuring} can not only evaluate the factual accuracy of LLM but also serve as a reliable calibration test, assessing the alignment between model confidence and factual accuracy. 
A well-calibrated LLM enables developers to judge the trustworthiness of their answers based on their stated confidence. 
Following the previous study, we conduct a calibration analysis by directly instructing the LLM to state its confidence after its answer when responding to a question (the prompt is seen in Appendix~\ref{sec:appendix2}).


As shown in Figure~\ref{figRQ3-1}, the results indicate that all LLMs' QA accuracy increases as their confidence increases. This suggests that users can 
select answers for which LLMs express higher confidence, potentially leading to more accurate outcomes. 
However, all evaluated LLMs exhibit a degree of overconfidence. Specifically, their performance falls well below the Y=X ideal calibration line, indicating that these LLMs tend to overestimate the accuracy of their answers. These findings indicate that LLM calibration still requires substantial improvement. 

\begin{center}
    \begin{myboxc}{\textbf{Summary: } The accuracy of LLMs shows a positive correlation with their stated confidence in Dev Knowledge QA. However, they tend to overestimate the accuracy of their answers. 
    }
    \end{myboxc}
\end{center}

\subsection{Capability Correlation Study}
\label{sec:cab_cor}

\begin{figure}[t]
    \centering
    \includegraphics[width=0.9\columnwidth]{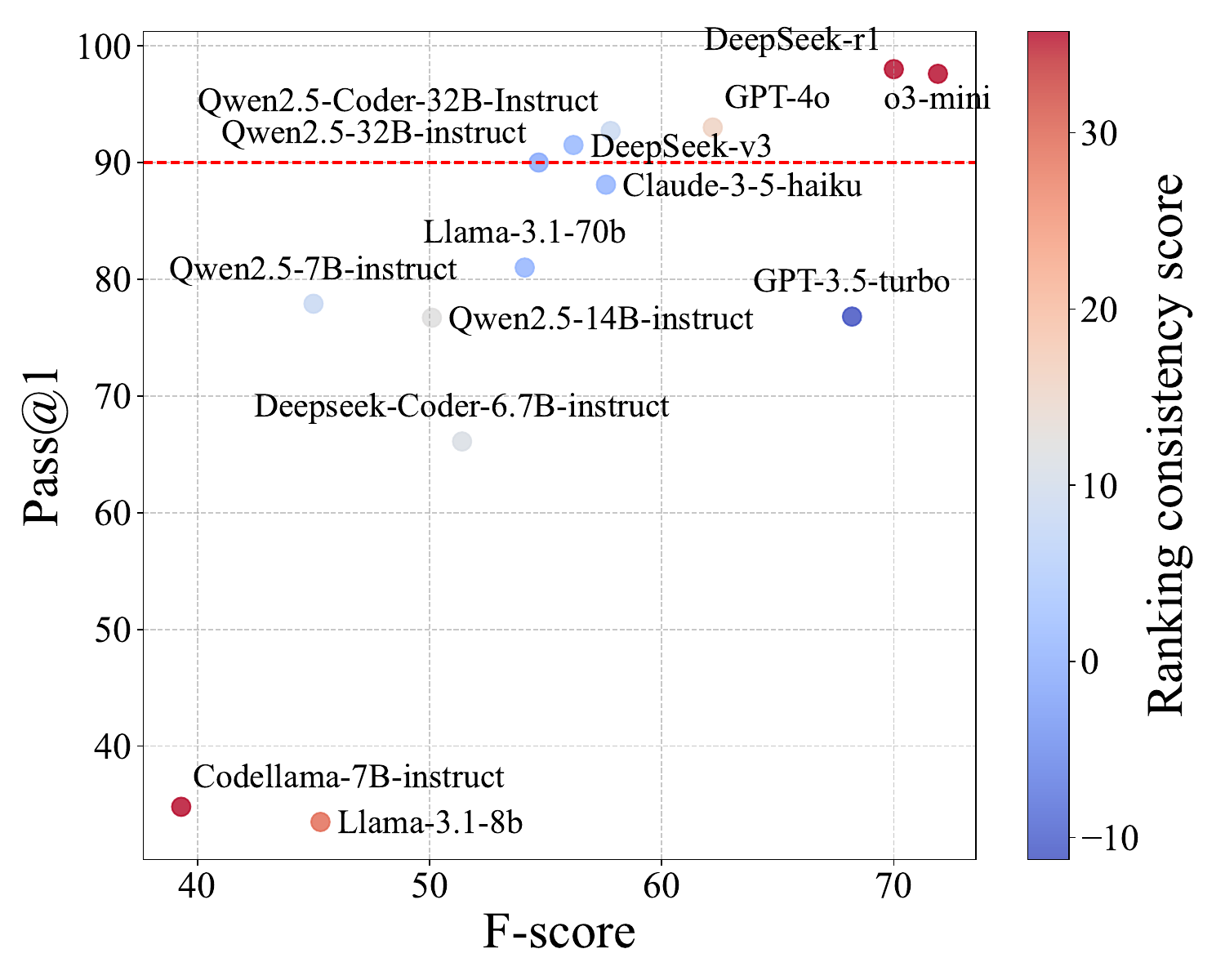}
    \caption{Results of different LLMs on SimpleDevQA and HumanEval.}
    \label{figRQ4}
\end{figure}

Based on our preliminary study in Section~\ref{sec:distribution}, code generation and Dev Knowledge QA are the most common tasks when developers interact with LLMs. To investigate how LLM performance relates across these two tasks, we compare evaluated LLMs on two benchmarks: HumanEval~\cite{chen2021evaluating} for code generation and SimpleDevQA for Dev Knowledge QA. We gather each model’s HumanEval pass@1\footnote{We collect pass@1 
from \url{https://evalplus.github.io/leaderboard.html}.} score alongside its SimpleDevQA F-score and plot these paired metrics in a scatter plot. 

Experimental results in Figure~\ref{figRQ4} demonstrate that LLMs that achieve higher pass@1 scores tend to also score higher on SimpleDevQA, indicating a strong relationship between code generation capability and development knowledge comprehension. 
Among them, o3-mini and DeepSeek-R1 exhibit outstanding performance on both datasets, demonstrating superior comprehensive capabilities. However, special cases exist. For example, Qwen2.5-7B-Instruct achieves nearly 80\% pass@1 on code generation benchmarks, yet its F-score is only about 45\%. This gap suggests that strong code generation ability does not guarantee equally strong performance on Dev Knowledge QA tasks.

\begin{center}
    \begin{myboxc}{\textbf{Summary: }Generally, LLMs with stronger code generation performance also exhibit stronger performance in Dev Knowledge QA, where o3-mini and DeepSeek-R1 show the best comprehensive performance. 
    }
    \end{myboxc}
\end{center}

\section{Conclusions}
 
In this paper, we propose SimpleDevQA, a multilingual Dev Knowledge QA benchmark built from real developer dialogues that covers broader development knowledge. 
Based on this dataset, we conduct extensive experiments. Experimental results show that first, closed-source models typically surpass open-source ones, and code LLMs generally outperform general LLMs of similar scale. Second, the Retrieval-Augmented Generation (RAG) strategy can enable smaller models to achieve performance comparable to larger models. Third, we find that LLMs’ accuracy generally increases with their stated confidence in SimpleDevQA. However, they tend to overestimate the accuracy of their answers. Finally, we find that LLMs with stronger code generation performance also exhibit stronger performance in the Dev Knowledge QA task.

By releasing SimpleDevQA to the community, we aim to help facilitate further research in software development. We hope that our benchmark will serve as a reliable tool for evaluating LLMs’ understanding of development knowledge.

\section*{Limitations}

We have identified the following limitations to our study.

\textbf{Limited benchmark size and data imbalance.} 
The current version contains only 5,529 instances across 10 domains and 3 languages, constrained by the limited size and coverage of the seed dataset during the data generation stage. Specifically, we sampled 3,000 real Dev Knowledge dialogues as the seed dataset, given the high cost of LLM inference and the labor-intensive manual verification required in the filtering stage. 
In addition, the dataset exhibits domain imbalance. Nearly half of the questions are in 53.7\% of the topics are
focused on "APIs \& Frameworks". Importantly, this imbalance is a direct consequence of our construction method: SimpleDevQA is inspired by real user-LLM dialogues, and thus the resulting distribution naturally reflects developers' knowledge-seeking tendencies and demand patterns in practice. In this sense, the skew is not merely a limitation, but also evidence of the benchmark's realism.
Nevertheless, the collection pipeline we provide is general for converting real-world dialogues into high-quality QA pairs. With this pipeline, additional data covering more languages and domains can be obtained. In future work, we plan to expand the seed dialogue pool to improve overall coverage and to examine whether scaling up real dialogue collection can naturally increase the volume of currently underrepresented categories, while still preserving the benchmark's real-world distributional characteristics.

\textbf{Potential bias from model generation and judgment.} 
The limitation is that the use of LLMs in both benchmark construction and evaluation may introduce bias. 
In benchmark construction, the QA pairs are generated by an LLM using the original dialogues and references as input. Although we apply a multi-stage filtering process to ensure correctness, this process may still introduce subtle biases, such as phrasing that reflects the LLM’s style rather than organic developer language. 
In evaluation, we set the temperature of the judge model = 0.5 and top-p = 1 to balance determinism and diversity. While this reduces the risk of extreme bias from greedy decoding, inherent randomness remains even after multiple runs, which may still affect the consistency, reproducibility, and introduce potential bias~\cite{ye2024justice, chen2024humans, zheng2023judging}. And we conducted a human evaluation and measured agreement with the LLM judge. The Cohen’s Kappa score~\cite{2012Interrater} between the LLM and human evaluation reaches 0.83 with an overall accuracy of 0.91. These results indicate a high level of agreement (Kappa > 0.8), supporting the high reliability of the LLM judge. 

\textbf{Limited scope of evaluated baselines.} While our current study provides comprehensive evaluations using zero-shot and RAG frameworks, the scope of our baseline comparisons remains somewhat restricted. Specifically, a limitation is the absence of baseline, such as deep research or agentic search. In the future, we will include deep research and agentic search as baselines to further understand the potential of our proposed benchmark.

\section*{Acknowledgments}

This work is supported by the National Natural Science Foundation of China (Grant No.92582202, No.92582110, No.62302534), and GMCC-SYSU Joint Lab for Smart Applications.



\bibliography{main}

\appendix

\section{Comparison of existing Dev Knowledge QA benchmarks}

The comparison between SimpleDevQA and other Dev Knowledge QA benchmarks is shown in Table~\ref{tabbenchmark}. We find that SimpleDevQA is not only grounded in real user queries but also covers diverse knowledge required throughout the software development process.

\begin{table*}[t]
  \centering
\small
\begin{tabular}{llll>{\centering\arraybackslash}p{2cm}  >{\centering\arraybackslash}p{2cm}}
\toprule
Benchmark   & Year & Size                             & Data Source              &  Built from Real User Queries & Diverse Dev Knowledge \\

\midrule

CodeQA~\cite{liu2021codeqa}      & 2021 & 190,000   & Code Snippets      & \XSolidBrush & \XSolidBrush                            \\
CS1QA~\cite{lee2022cs1qa}       & 2022 & 9,237     & Textbooks, Edu Materials & \Checkmark & \XSolidBrush                          \\

RepoQA~\cite{liu2024repoqa}  & 2024 & 500 & Real Code Repositories   & \XSolidBrush &  \XSolidBrush \\
InfiBench~\cite{li2024infibench} & 2024 & 234 & Real questions & \Checkmark &  \XSolidBrush \\ 
CoReQA~\cite{chen2025coreqauncoveringpotentialslanguage} & 2025 & 1,563 & Real issues and comments & \Checkmark &  \XSolidBrush \\

\midrule
SimpleDevQA (ours) & 2025 & 5,529    & Real User Dialogues, Web   & \Checkmark & \Checkmark \\   
\bottomrule
\end{tabular}
\caption{Comparison of existing Dev Knowledge QA benchmarks.}
\label{tabbenchmark}
\end{table*}

\section{Extraction of the software development-related dialogues}
\label{sec:appendix_extract}

\begin{figure}[t]
    \centering
    \includegraphics[width=\columnwidth]{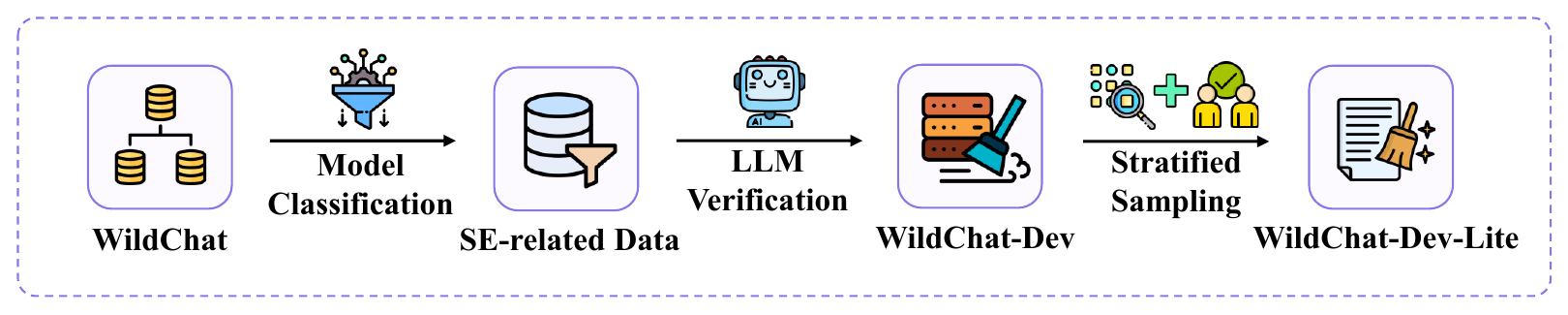} 
    \caption{Data processing workflow for extracting software development–related conversations.}

    \label{figPS} 
\end{figure}

Following WildChat~\cite{zhao2024wildchat},  we use a prompt-classification model~\cite{valpy2024promptclassification} to filter the dataset and obtain 376,888 dialogues related to software engineering. 
To further enhance data quality, we employ Llama3-Instruct-8B~\cite{meta_llama3_8b_instruct} to process the dialogues, filtering out conversations irrelevant to software development. This step results in the WildChat-Dev dataset, which includes 103,112 dialogues focused on development.
To facilitate our manual annotation and analysis of real dialogues, we apply stratified sampling to 103,112 dialogues based on user language, dialogue length, and the interacted model within the data, which results in the WildChat-Dev-Lite dataset.  To compute the sample size, we follow previous studies~\cite{yang2023definition,11030227}, use the random sampling method based on the confidence interval~\cite{wikipedia:confidence_interval}. We set a confidence interval of 1 and a confidence level of 95\%, and compute that the sample size is 8,786~\cite{surveysystem_sscalc_2023}. 
At the end, we obtain the WildChat-Dev-Lite dataset, which includes 8,786 real dialogues. The whole data processing workflow is shown in Figure~\ref{figPS}.



\section{Classification of the real dialogues}
\label{sec:appendix_classfy}

To analyze the distribution of software engineering–related tasks in real-world dialogues, we apply the following pipeline to categorize real-world dialogues. First, we manually filter dialogues from WildChat-Dev-Lite that are unrelated to software development. Then, we use the open card sorting method~\cite{lewis2010open} to summarize the tasks involved in the dialogues from WildChat-Dev-Lite, and the final eight categories are inspired by prior works~\cite{zhang2023unifying, hou2024large, niu2022deep}, as shown in Table~\ref{t2}.
Finally, we manually classify the remaining dialogues into 8 task categories. Specifically, we first conduct a pilot study in which three annotators jointly label a small sample of dialogues and discuss the criteria. During the annotation process, two annotators independently label each dialogue. They then cross-check the results and resolve discrepancies through discussion. If no agreement can be reached, a third annotator is involved, and final decisions are made via majority voting.
To assess annotation reliability, we calculate Krippendorff’s alpha~\cite{2011Computing}, which is 0.908, indicating strong agreement. The final average annotation accuracy is 0.924.

\begin{table}[t]
\centering
\resizebox{\columnwidth}{!}{%
\begin{tabular}{ll}
\toprule
\textbf{Task Category} & \textbf{Task Description} \\
\midrule
Dev Knowledge QA & QA that questions knowledge during development \\
Code Generation  & Generating code examples based on requirements \\
Code Debugging   & Identifying errors in existing code \\
Program Repair   & Fixing defective or buggy code \\
Code Translation & Converting code from source to target language \\
Code Editing     & Modifying or enhancing functionality of existing code \\
Comment Generation & Generating explanatory comments for code \\
Test Generation  & Creating test cases for code \\
\bottomrule
\end{tabular}%
}
\caption{Description of software engineering-related tasks in real scenarios}
\label{t2}
\end{table}

\section{Generation of QA pairs}
\label{sec:appendix1}

\begin{figure}[t]
    \centering
    \includegraphics[width=\columnwidth]{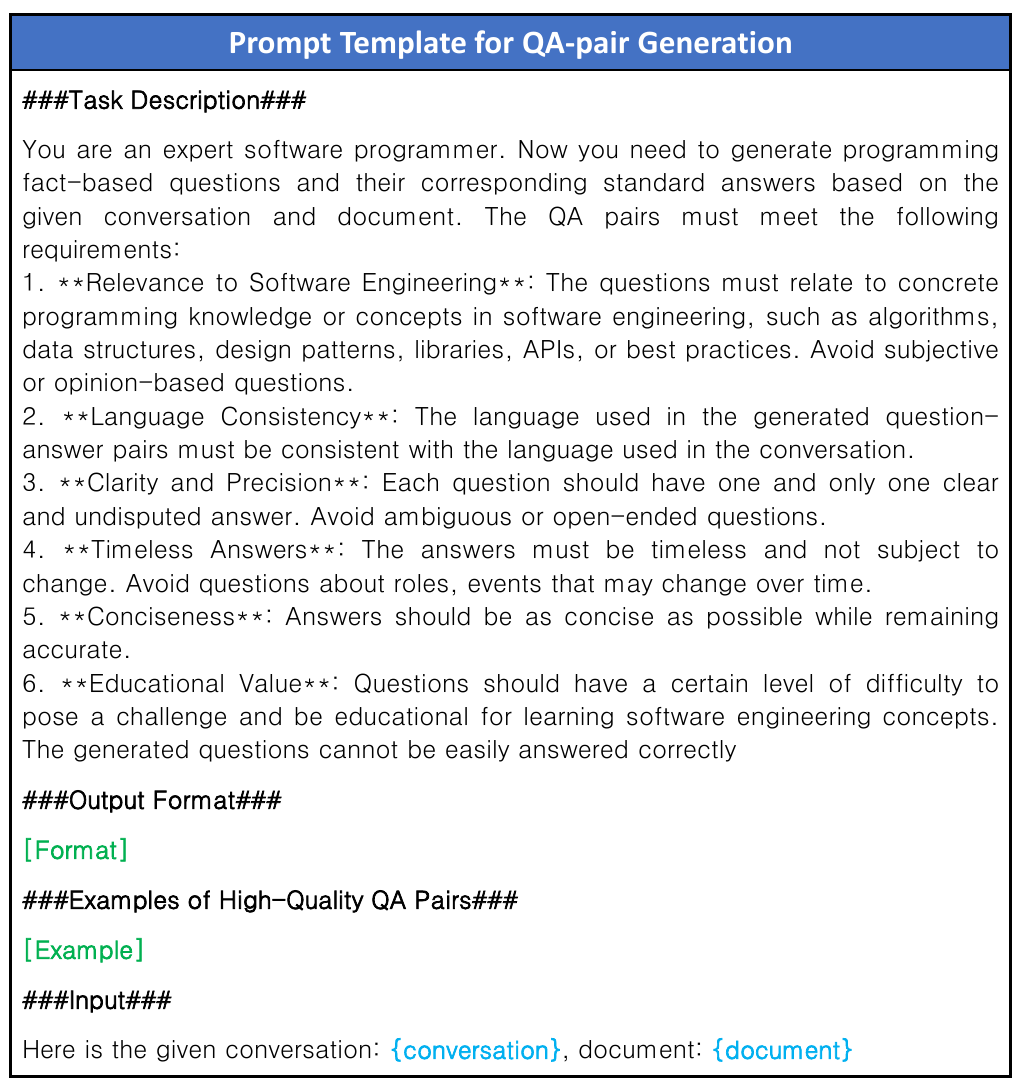} 
    \figmargin
    \caption{The prompt template for generating QA pairs.}
    \label{prompt} 
\end{figure}

We use GPT-4o~\cite{openai2024gpt4o} to generate Dev Knowledge QA pairs. The prompts for generating QA pairs are presented in Figure~\ref{prompt}. Specifically, we request that LLM generate 3 QA pairs for one dialogue at the same time. Besides, we provide 10 demonstrations to enhance the quality of generated Dev Knowledge QA pairs from LLMs.

\begin{figure}[t]
    \centering
    \includegraphics[width=\columnwidth]{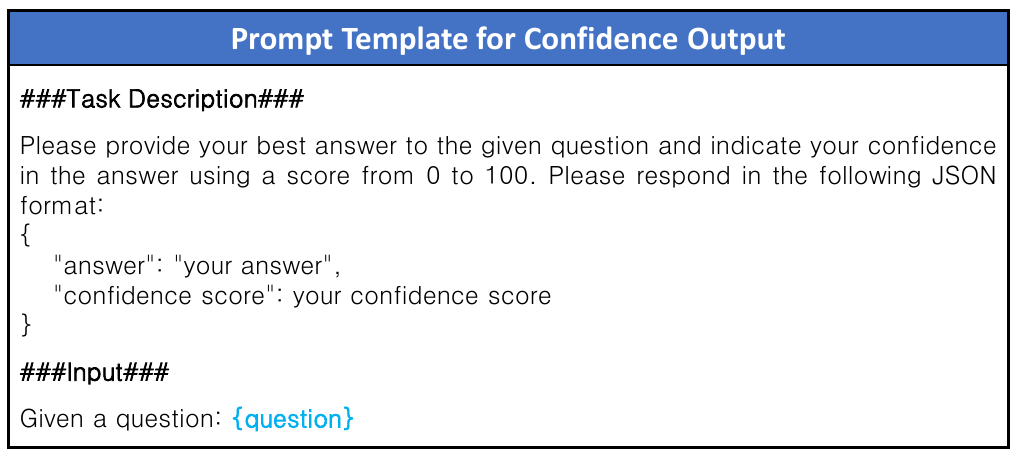} 
    \caption{The prompt for guiding LLM to output confidence.}
    \label{prompt2} 
\end{figure}


\section{Characteristics of SimpleDevQA}
\label{sec:appendix3}

\begin{table*}[t]
    \centering
    \vspace{-0.5em}

    \begin{minipage}[t]{0.68\linewidth}
        \centering
        \subcaption{Knowledge Domain.}
        \resizebox{\linewidth}{!}{
            \begin{tabular}{ccc}
                \toprule
                \textbf{Category}                     & \textbf{Domain}                       & \textbf{Size} \\
                \midrule
                \multirow{2}{*}{Syntactic knowledge}  & APIs \& Frameworks                   & 2967 \\
                                                      & Programming language syntax          & 1168  \\
                \midrule
                \multirow{7}{*}{Semantic knowledge} & Software development \& engineering  & 495   \\
                & Algorithms \& Data structures        & 359  \\
                                                      
                                                      & Database management \& SQL           & 224   \\
                                                      & Computer organization \& architecture& 147   \\
                                                      & Object-oriented programming          & 83   \\
                                                      & System design               & 78   \\
                                                      & Compiler design                 & 5    \\
                                                      & Computer networks                & 3    \\
                \bottomrule
            \end{tabular}
        }
    \end{minipage}%
    \hfill
    \begin{minipage}[t]{0.3\linewidth}
        \centering
        \subcaption{Language Usage.}
        \resizebox{0.7\linewidth}{!}{
        \small
            \begin{tabular}{cc}
                \toprule
                \textbf{Language} & \textbf{Ratio} \\
                \midrule
                English  & 36.75\% \\
                Chinese  & 37.46\% \\
                Russian  & 25.79\% \\
                \bottomrule
            \end{tabular}
        }

        \vspace{1em} 

        \subcaption{Average Token Length.}
        \resizebox{0.6\linewidth}{!}{
        \small
            \begin{tabular}{cc}
                \toprule
                \textbf{QA}        & \textbf{Length} \\
                \midrule
                Question  & 27.74  \\
                Answer    & 7.99   \\
                \bottomrule
            \end{tabular}
        }
    \end{minipage}
    \caption{\small SimpleDevQA Statistics.}
    \label{tab:benchmark-statistics}
\end{table*}

SimpleDevQA covers the three most prevalent languages from the WildChat-Dev-Lite dataset: English, Chinese, and Russian. 
Specifically, it includes 2,032 English QA pairs, 2,071 Chinese QA pairs, and 1,426 Russian QA pairs.

Additionally, questions in our benchmark have a length of 27.74 tokens
\footnote{\url{https://github.com/openai/tiktoken}.}
, while answers have a length of 7.99 tokens on average, as shown in Table~\ref{tab:benchmark-statistics}. In addition to these, each QA pair is also accompanied by multiple web-retrieved references, which include corresponding webpage URLs and text snippets that can be used to verify the factual accuracy of the answers.

To categorize the knowledge domains covered by the QA pairs in the dataset, we manually classify the questions using a taxonomy adapted from a previous study~\cite{manh2024codemmlu}. For ambiguous or cross-category queries, annotators were instructed to prioritize the category that is most central to answering the question, while recording secondary categories when applicable for internal auditing. The classification includes:
\begin{itemize}[leftmargin=10pt]
   \item\textbf{Syntactic questions: }These focus on programming language grammar and API usage, such as language-specific functions or common library usage, etc.
   \item\textbf{Semantic questions: }These target more abstract programming concepts, such as algorithms, data structures, and object-oriented principles, etc.
\end{itemize}

The classification results, shown in Table~\ref{tab:benchmark-statistics}, indicate that the benchmark contains 4,135 syntactic questions and 1,394 semantic questions, covering 10 distinct development knowledge domains.



\section{Analysis of stated confidence}
\label{sec:appendix2}

We instruct the LLM to state its confidence after its answer when responding to a question for the specific prompt, as shown in Figure~\ref{prompt2}. 

\section{Case Studies}
\label{sec:appendixcase}

\begin{figure}[t]
    \centering
    \includegraphics[width=\columnwidth]{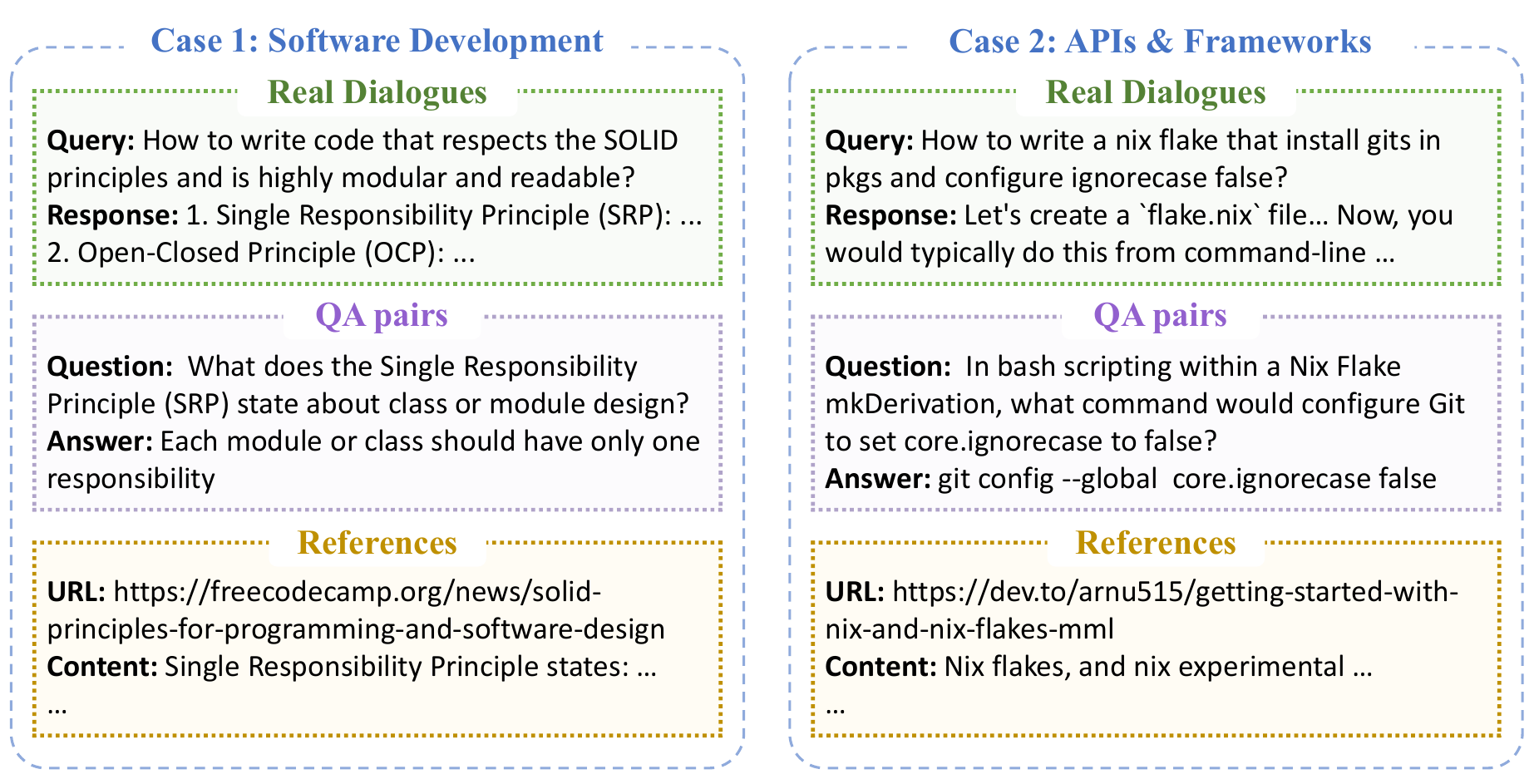} 
    \caption{
    Case studies in SimpleDevQA.
    }
    \label{fig-morecase} 
\end{figure}

In our dataset, we categorize and analyze cases in SimpleDevQA. 
We identify the following aspects that highlight the advantages of SimpleDevQA.

\textbf{Simple and Accurate Evaluation.} 
A key challenge in using Dev Knowledge QA derived from real-world dialogues is that the resulting QA pairs are often \emph{hard to evaluate reliably}. As shown in Figure~\ref{fig-morecase}, real user queries are frequently open-ended (e.g., “How to write code that respects SOLID…”), and the corresponding responses may mix subjective guidance with multiple factual statements and multiple valid solution paths. 
Consequently, even when a model produces a semantically correct response, it may differ substantially in wording, structure, or chosen solution route, making automated evaluation unreliable.
SimpleDevQA mitigates this issue by focusing on knowledge-seeking questions with short, unambiguous, and verifiable answers through a multi-step curation process. 
Moreover, each QA pair is accompanied by a supporting reference, which provides explicit evidence for adjudication and further improves the accuracy and consistency of LLM-as-judge evaluation.

\textbf{Broad Development Knowledge Coverage.} 
SimpleDevQA is designed to reflect development knowledge needs beyond code-centric understanding. For instance, Figure~\ref{fig-morecase} includes questions about software engineering principles (e.g., SRP under the SOLID principles) 
and practical workflow knowledge, such as configuring Git via terminal commands in real development environments. 
These examples cover \textit{Software Development \& Engineering} (software principles, version control practices) and \textit{APIs \& Frameworks}, demonstrating that the benchmark evaluates the broader knowledge developers routinely rely on during end-to-end software development.

\end{document}